\documentclass[12pt]{iopart}
\usepackage{fullpage}
\usepackage{amsmath}
\usepackage{comment}
\usepackage{hyperref}
\usepackage[pdftex]{graphicx}
\usepackage[margin=1in]{geometry}
\usepackage[margin=1em,font=scriptsize,labelsep=period]{caption}

\usepackage[backend=biber, sorting=none]{biblatex}
\usepackage{listings}
\usepackage{xcolor}

\definecolor{codegreen}{rgb}{0,0.6,0}
\definecolor{codegray}{rgb}{0.5,0.5,0.5}
\definecolor{codepurple}{rgb}{0.58,0,0.82}
\definecolor{backcolour}{rgb}{0.95,0.95,0.92}

\lstdefinestyle{mystyle}{
    backgroundcolor=\color{backcolour},   
    commentstyle=\color{codegreen},
    keywordstyle=\color{magenta},
    numberstyle=\tiny\color{codegray},
    stringstyle=\color{codepurple},
    basicstyle=\ttfamily\footnotesize,
    breakatwhitespace=false,         
    breaklines=true,                 
    captionpos=b,                    
    keepspaces=true,                 
    numbers=left,                    
    numbersep=5pt,                  
    showspaces=false,                
    showstringspaces=false,
    showtabs=false,                  
    tabsize=2
}

\begin{document}
\title{Compression and information entropy of binary strings from the collision history of three hard balls}
\author{M. Vedak$^{1,2}$, and G. J. Ackland$^2$}
\address{$^1$University of Vienna, 1010 Vienna, Austria 

$^2$School of Physics and Astronomy, University of Edinburgh, Edinburgh EH9 3FD, United Kingdom}
\ead{matejvedak@gmail.com, gjackland@ed.ac.uk}
\vspace{10pt}
\begin{indented}
\item[]October 2022
\end{indented}

\begin{abstract}
We investigate how to measure and define the entropy of a simple chaotic system, three hard spheres on a ring.  A novel approach is presented, which does not assume the ergodic hypothesis. It consists of transforming the particles' collision history into a sequence of binary digits.  We then investigate three approaches which should demonstrate the non-randomness of these collision-generated strings compared with random number generator created strings:  
         Shannon entropy, diehard randomness tests and compression percentage. We show that the Shannon information entropy is unable to distinguish random from deterministic strings. The Diehard test performs better, but for certain mass-ratios the collision-generated strings are misidentified as random with high confidence. The zlib and bz2 compression algorithms are efficient at detecting non-randomness and low information content, with compression efficiencies that tend to 100\% in the limit of infinite strings. Thus "compression algorithm entropy" is non-extensive for this chaotic system, in marked contrast to the extensive entropy determined from phase-space integrals by assuming ergodicity.
\end{abstract}

\submitto{Journal of Physics Communications}

\section{Introduction}

The validity of statistical physics is determined by its central assumption: a physical system must sample a representative area of its theoretical phase space. This leads to the question of the {\it ergodic hypothesis}: Points in a moving ergodic system will eventually visit all parts of the phase space that the system moves in. With respect to physical systems, ergodicity is the assumption which means that thermodynamics properties can be calculated from phase-space integrals.
Ergodicity implies that the average behaviour of the system can be deduced from the time-averaged trajectory of a 'typical' point. It is further assumed that sampling over a long time series is equivalent to sampling over a large phase space.  Equivalently, ergodicity means that taking a sufficient amount of random samples of the process can represent the average statistical properties of the whole process. Many of the predictions of ergodic theory are proven for hard sphere systems, however, strict ergodicity is a sufficient {\it but not essential} condition for this:
it is not clear that a deterministic system will actually visit all available microstates \cite{Vranas1998} \cite{Zheng1996}. 

Entropy is typically described in terms of the information content of a distribution, and can be measured in terms of either probabilities, or by counting the number of microstates in a macrostate.  For a system exhibiting deterministic chaos, the true information content is low - everything is fully determined by the initial conditions.  By contrast, the ergodic hypothesis is that the system will fully explore its phase space, so measures based on either probability or counting microstates suggest high entropy.  

This is why explorations of ergodicity in relatively simple systems became an important area of research. It is of interest to find systems that are complex enough to exhibit non-trivial ergodicity but simple enough so we can actively examine the phase space in which they move. Many such systems have been explored, and of particular importance are hard-ball systems because of their simplicity (for example, \cite{Simányi2004}).

Another related concept is "stochastic simulation" - in a computer simulation this can be defined as any trajectory which involves random numbers, e.g. Langevin dynamics. This is distinct from deterministic chaos, for which a simulation maps to Newtonian mechanics and does not use a random number generator. It is possible to use chaotic or stochastic dynamics to explore the same  phase space\cite{Cox2000}.

We investigate one of the simplest systems exhibiting deterministic chaos: three pointlike particles that move on a 1D periodic ring. This specific system was first investigated by one of us 20 years ago \cite{ackland1993} as an example of the more general issue of how ergodicity, chaos and integrability allow statistical mechanic models to represent thermodynamic problems\cite{fermi1955studies,berman2005fermi}.
We have previously shown \cite{ackland1993,Cox2000} that there are
periodic trajectories in this system, such that some regions of the phase space cannot be accessed.
The phase space of this system is six dimensional, with axes consisting of three positions and three momenta of the particles. Conservation of energy and momentum reduce the accessible region to a four-dimensional hypersurface within this six-dimensional phase space. Particles continuously explore position space, while coordinates in momentum space remain constant until a collision occurs, after which they take on new constant values. Labelling the initial momenta with $p_1$ and $p_2$ and the subsequent post collision momenta with $q_1$ and $q_2$ of particles with masses $m_1$ and $m_2$ the update formulae are:
\begin{align}
    q_1 &= \frac{m_1 - m_2}{m_1 + m_2}p_1 + \frac{2m_1}{m_1 + m_2}p_2, \\
    q_2 &= \frac{2m_2}{m_1 + m_2}p_1 + \frac{m_2 - m_1}{m_1 + m_2}p_2.
\end{align}
The outcome of every collision can be represented by the operation of a matrix on a vector of momenta $(p_1, p_2, p_3)$, a state vector in the phase space. Any sequence of collisions can be mapped onto a product of such matrices \cite{ackland1993}.

The spatial probability distribution was found to be triangular  for each of the three particles, going to zero where a triple collision occurs (Fig, \ref{fig:phasedistribution}). The analytical derivation of this result can be found here \cite{Li2003alt}. Momentum space distributions assume a vastly different shape. As outlined in \cite{ackland1993}, the shape of three particle momentum distributions is an arcsine distribution:
\begin{equation} P(p_i) \propto (p_{i,max}^2 - p_i^2)^{-1/2}. \end{equation}
This distribution is the projection of an ellipse onto one axis in the 3D momentum phase space (the ellipse being the region that satisfies the conservation of energy and momentum). The mean energy of each particle can be calculated \cite{Cox2000}
\begin{equation} E_i = \int \frac{p_i^2}{2m_i} P(p_i)dp_i = \frac{M-m_i}{2M} E_{tot}. \label{eq:equi}\end{equation}
It is proportional to the mass of the other two particles. Departure from the usual equipartition result is due to the fact that the system is in the 'molecular-dynamics' ensemble of statistical mechanics where energy, volume, and the number of particles are conserved, and the total momentum is zero. \cite{Tahir1988}.  A generalized version for any number N of particles can be found \cite{ackland1993}:
$$ \langle E_i \rangle = \frac{\sum_j m_j - m_i}{\sum_j m_j} \frac{E}{N-1}. $$
which satisfies equipartition in the high-N limit. Distributions of particle positions converge quickly to Gaussianity as the number of particles is increased.  Similar results with non-trivial patterns are found in stochastic models in higher dimension\cite{vidgop2011evolution,vidgop2014emergence}.  

This system has been shown to be equivalent to a billiard problem in a triangular stadium \parencite{Glashow1997}. With the triangular system in mind, it is easy to see that there are periodic orbits (example of a right triangle where the billiard bounces off the two non-hypotenuse sides at a right angle). Since one cannot escape the periodic orbit, and the dynamics are time-reversible, one cannot enter it either. In this way, the microstates of a periodic orbit remain inaccessible to the general case, which violates strict ergodicity.
Explorations of whether non-ergodicity invalidates statistical mechanics have already been made \parencite{Cox2000}, where it was shown that this chaotic system samples the phase space efficiently, such that thermodynamic integrals converge to the same value as if it was ergodic. This paper focuses on information and entropy. We generate binary strings based on particle collisions and compare those strings to completely randomly generated ones. The more random the collision generated strings end up being, the stronger belief in the ergodicity hypothesis can be held and the system is better at exploring its phase space. To the best of our knowledge, this is a novel approach that has not been explored yet.

\section{Methods}

Start with three particles. Sorting by their starting position, we index them: $0, 1, 2$. Meaning at the onset of the simulation $x_0 < x_1 < x_2$ ($x_i \in [0, 1\rangle$). Collisions are simulated and say that the following collisions happen: particles 1 and 2 collide, particles 0 and 1 collide, particles 0 and 2 collide, particles 0 and 1 collide. In a simple notation, each collision is represented by the particle which does not collide, so those collisions would be written down as follows: 0212.  However,
after two particles collide there are only two options, 
so we investigated two encoding methods (EM1 and EM2) to convert this ternary string into a binary string.
\begin{enumerate}
    \item EM1
After a collision, either the 'left' particle collides next with the third one, or the 'right' particle does. The trajectory is then mapped to a binary string in the following way:  A 'left' particle collision is encoded by a 0, collision of a 'right' particle is encoded as 1. Using the above collisions as an example, we will explicitly write out the generated string. The first collision is between particles 1 and 2, and subsequently, particle 1 (on the 'left' of the collision) collides with the third particle, particle 0. At that point, the string we are generating will look like: \\
${} \hspace{7cm} s = 0$. \\

The subsequent collision is between particles 2 and 0; again, 0 was the particle on the 'left' of the previous collision so the string is updated accordingly: \\
${} \hspace{7cm} s = 00$. \\

The final collision in the example is $(0, 1)$, where 0 was the particle on the 'right' in the previous collision, so a 1 is appended to the string: \\
${} \hspace{7cm} s = 001$. \\

\item ME2: Each entry in the ternary string is replaced by 1 is both adjacent entries are the same, or zero is they are different.  So e.g. 0212 becomes .01. with the dot denoting entries which would require an extended string to generate
\end{enumerate}

These encoding produce different binary string representing the same sequence. They are referred to as collision generated strings, as opposed to random strings,  created using random number generators (RNG, we used python's numpy.random module, so they are technically pseudo-random numbers).

Perhaps the most obvious first test to conduct is the inspection of entropy of the strings at hand. The information theory approach was used, and Shannon entropy was utilized:
$$ H(X) = - \sum_{i=1}^2 P(x_i) \log_2 (P(x_i)). $$
In the binary string case, X is the binary string, $P(x_1)$ corresponds to the total proportion of 0s in the string, while $P(x_2)$ corresponds to the total proportion of 1s in the string.  As shown in figure \ref{fig:unified_comparison}, the system maximises its Shannon entropy by equalising the number of "left" and "right" collisions. Thus Shannon entropy is unable to distinguish between low-information deterministic trajectories, and high-information random trajectories.

Furthermore, there are many other ways of randomness testing. In this context, we are interested in statistical randomness. A sequence of characters is statistically random if it does not contain any recognizable patterns or regularities. It is important to note that statistical randomness does not necessarily imply true randomness or objective unpredictability. This is due to the fact that any inspected and tested sequence must be finite. Any tests conducted on such a finite sequence can only prove 'local' randomness, even though we are interested in the true, 'global'.

The first randomness tests were published by M. G. Kendal and Bernard Babington Smith \cite{Kendall1938} and originally consisted of four tests:
\begin{itemize}
    \item The frequency test checked whether the number of 0s, 1s, 2s, and other digits was roughly the same.
    \item The serial test did the same thing as the frequency test, but for sequences of two digits at a time, comparing the observed frequencies with the hypothetical predictions.
    \item The poker test tested for specific sequences of five numbers at a time based on hands in the game of poker.
    \item The gap test looked at the distance between zeroes (00 is a distance of 0, for example).
\end{itemize}
Since then, randomness tests have increased in complexity and have improved. Perhaps the most famous and popular are the so-called diehard tests \cite{Marsaglia1995}, a battery of statistical tests that measure the quality of a random number generator \cite{bellamy2013}.

Another way of testing how random a sequence of characters is is by using a compression algorithm and seeing how compressed the data turns out to be. Furthermore, comparing the compression percentage of collision generated strings with actual RNG strings provides with a meaningful comparison of randomness. This works because the compression algorithms themselves try to exploit repetitiveness and correlations of the given data, and therefore they will create better compression on non-random data sequences.

Here, we used one diehard test - the Wald-Wolfowitz runs test \cite{NIST}. It can be used to test the hypothesis that the elements of the sequence are mutually independent. It uses runs - a run is defined as a series of increasing values or a series of decreasing values. For example, the following sequence "0010" consists of two runs - there is a single change from "0" to "1" and a single value change from "1" to "0". The number of increasing or decreasing values is the length of the run. In a random data set, the probability that the next value is smaller or larger follows a binomial distribution, a fact that forms the basis of the runs test. The test's null hypothesis is that the sequence was produced in a random manner. Define the following values: $n_1$ is the number of positive (increasing) runs, $n_2$ is the number of negative (decreasing) runs. The expected number of runs is then given by
$$ \overline{R} = \frac{2n_1n_2}{n_1+n_2} + 1,$$
the standard deviation by
$$ s_R^2 = \frac{2n_1n_2(2n_1n_2 - n_1 - n_2)}{(n_1 + n_2)^2(n_1 + n_2 - 1)}. $$
The test statistic is
$$ Z = \frac{R - \overline{R}}{s_R} $$
where R is the observed number of runs. The runs test rejects the null hypothesis if $|Z| > Z_{1-\alpha/2}$ where $\alpha$ is the desired significance level. For example, if we desire a 5\% significance level, the test statistic with $|Z|$ greater than $1.96$ indicates non-randomness.

Finally, we use compression algorithms to detect information content: a truly random string is incompressible. Algorithms we used are zlib \cite{zlib_github}, bzip2 \cite{bzip2_github} and a custom run-length encoding algorithm. Zlib and bzip2 are very complex algorithms and we will not go into detail about their inner workings. The custom algorithm works by collecting subsequent occurrences of same characters into a number of occurrences and the character itself. It simply replaces same characters that occur in a series with the number of occurrences and the character itself. For example, the string $aabaaabbb$ is compressed to $2a1b3a3b$.

\section{Results \& Discussion}

\begin{figure}[h!]
    \centering
    \includegraphics[width=0.8\textwidth]{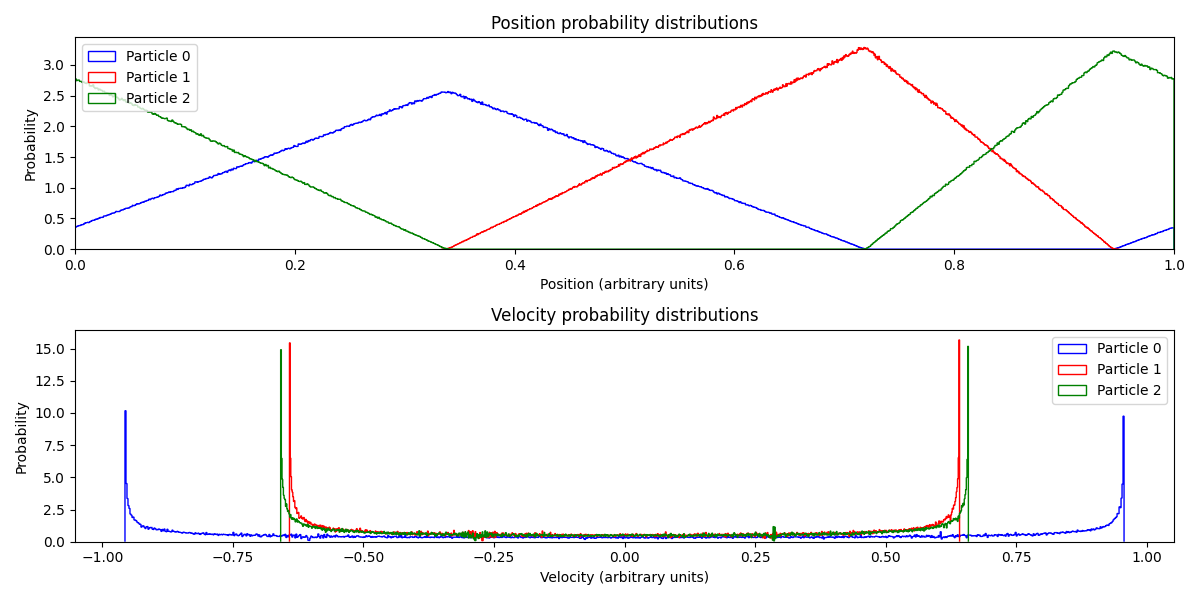}
    \caption{Position and velocity probability distributions for three particles on a ring. Random initial conditions. Position probability distributions follow the triangular distribution and each particle is confined to its region, restricted by points where a triple collision occurs. Velocity probability distributions are of the form $P(v_i) \propto \frac{1}{\sqrt{v_{i, max}^2 - v_i^2}}$ where $v_{i,max}$ is determined by the masses of the particles and the total energy of the system.
    \label{fig:phasedistribution}}
\end{figure}

We start with the verification of the previously mentioned results for a system of three particles on a ring. 
Example position and velocity distributions can be seen in figure \ref{fig:phasedistribution}. The triangular distribution in position space and the arcsine distribution in velocity space are nicely depicted.

\begin{figure}[h!]
    \centering
    \includegraphics[width=1\textwidth]{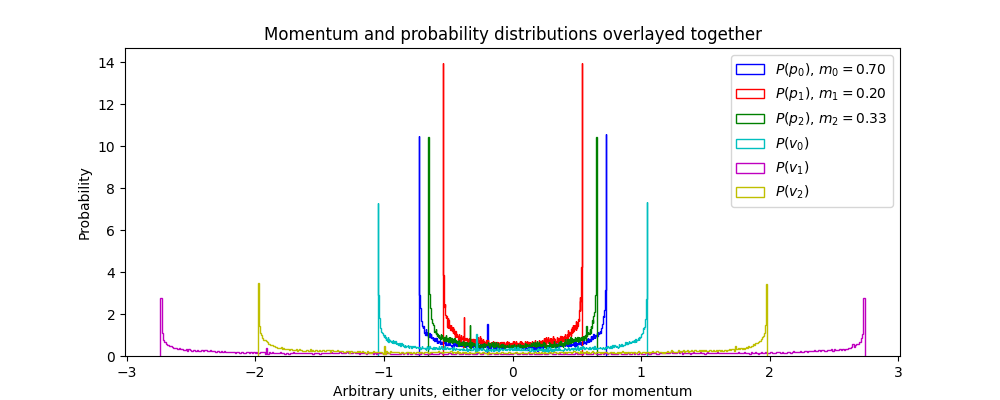}
    \caption{Sample momentum and velocity distribution shown on the same plot. Typical distribution results can be observed - small mass particles have narrow momentum distributions and broad velocity distributions, high mass particles have wide momentum distributions and narrow velocity distributions.}
    \label{fig:mom_and_vel}
\end{figure}

Interesting to note that, following equation \ref{eq:equi}, the smaller the mass of the particle, the wider its velocity distribution will be, but its momentum distribution will be less spread out. In figure \ref{fig:mom_and_vel}, we can see that the particle with the smallest mass has the  narrowest momentum distribution but the widest velocity distribution. Due to momentum conservation, lighter particles must acquire higher velocities to offset more massive particles' momentum.

\begin{figure}[h!]
    \centering
    \includegraphics[width=1\textwidth]{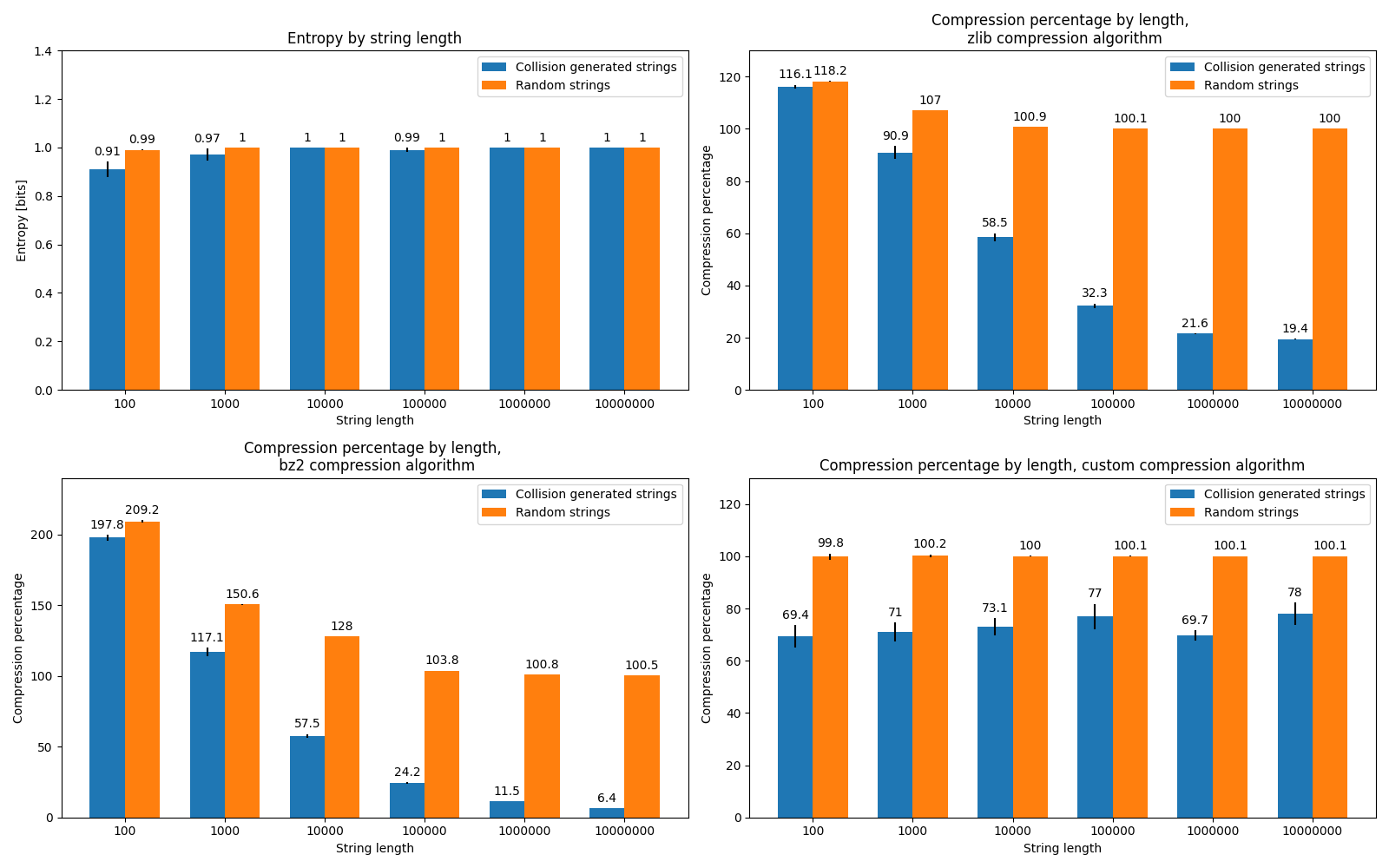}
    \caption{Performance of the chosen algorithms on both collision generated strings from EM1 (blue) and random generated (orange) strings. Top left is the Shannon entropy comparison, top right is the zlib algorithm compression percentage, bottom left is the bz2 algorithm compression percentage, and bottom right is the custom compression algorithm compression percentage (example compression: $aabccc \xrightarrow{} 2a1b3c$).  Results are averaged over 40 runs of randomized initial starting conditions for positions, momenta and masses }
    \label{fig:unified_comparison}
\end{figure}

\begin{figure}[h!]
    \centering
    \includegraphics[width=1\textwidth]{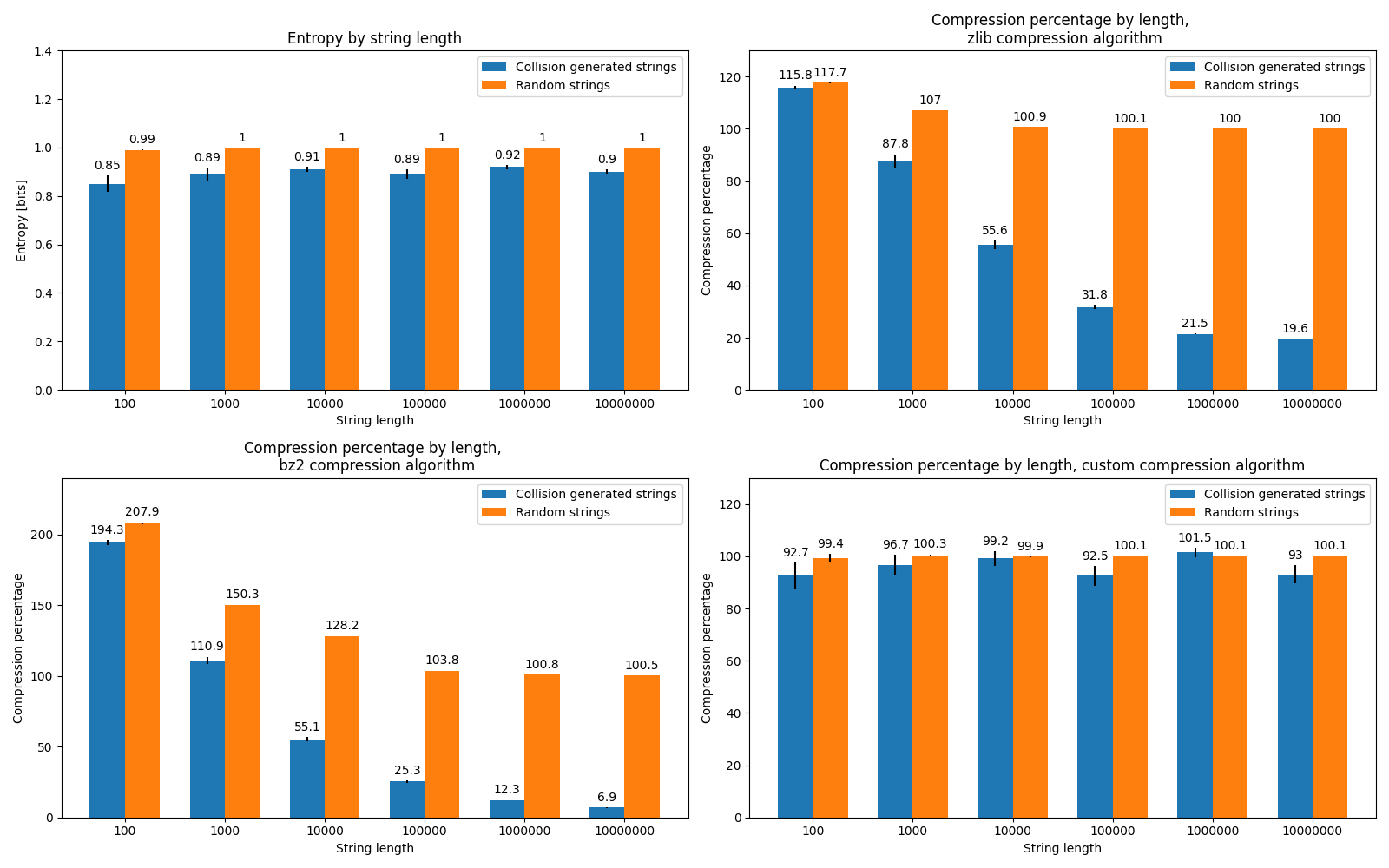}
    \caption{As fig \ref{fig:unified_comparison} With EM2 collision genererated strings}
    \label{fig:unified_comparison2}
\end{figure}

Moving on to string generation, Figures \ref{fig:unified_comparison} and \ref{fig:unified_comparison2} show the information entropy and compression percentages of strings with differing lengths.  Immediately, differences can be observed. The Shannon entropy of  random strings quickly take on the value of 1 bit of entropy ($H(x)=\ln2)$, while the collision generated strings only converge at longer lengths, to 1 for EM1, and $<1$ for EM2

A similar pattern is seen in the custom run-length encoding algorithm, where collision strings are compressed to $\sim 70\%$ (EM1) but only $\sim 93\%$ (EM2) while random strings see no compression at all. However, in this case string length has no substantial effect on the compression percentages. 

The zlib and bz2 algorithms exploit these long runs of identical digits and other complex methods to achieve compression. They compress collision generated strings substantially better than random string, and become increasingly effective for large strings. Notably, they are equally good at compressing EM1 and EM2 strings.

\begin{figure}[ht]
    \centering
    \includegraphics[width=1\textwidth]{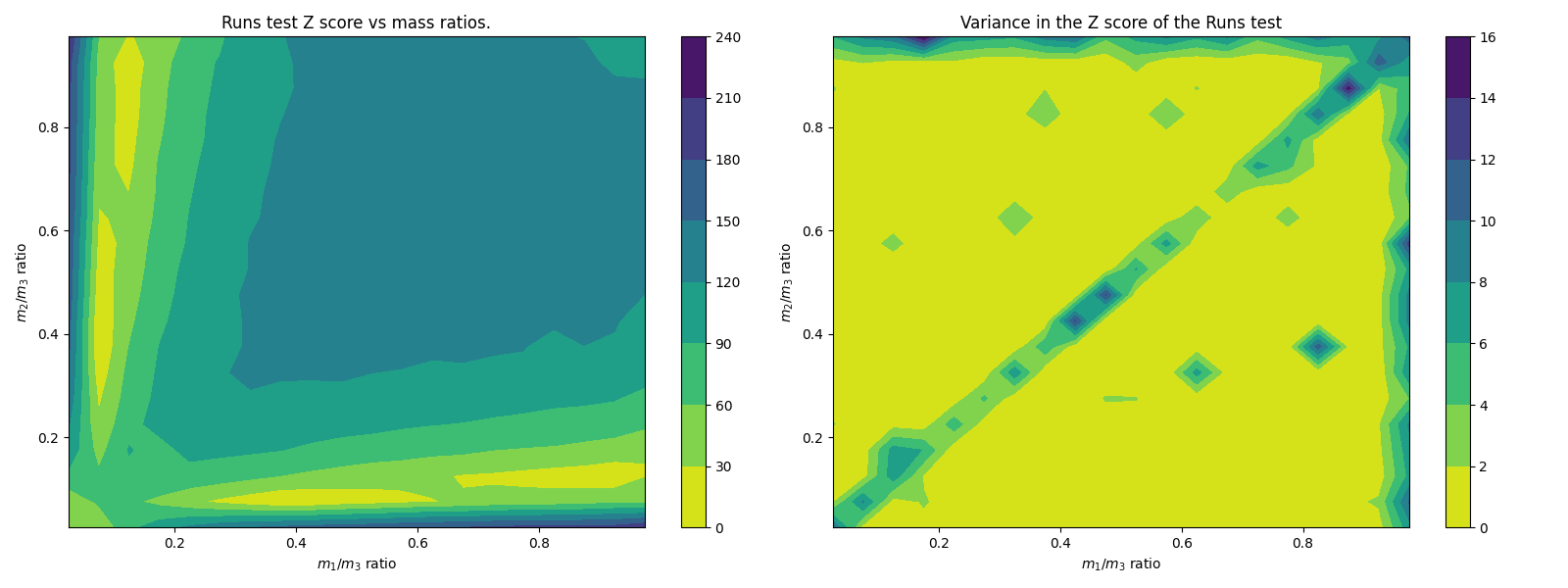}
    \caption{Results of the Wald-Wolfowitz runs test on collision generated strings with various mass ratios. For each combination of mass ratios, ten strings were produced from randomized initial positions and velocities, the results were averaged and shown in the figure. On the left, we see that most of the input space produces relatively high Z scores. There are a few valleys of low Z scores. Variance in the Z score is reasonably low, barring a few areas with close to two equal masses, enforcing the belief that mass ratios rather than initial velocities 'control' the randomness of the generated strings.
    The mass ratios were varied for a total of 20 equally spaced datapoints, ranging from $0.01$ to $1$. For each mass ratio combination, ten $100,000$-long strings were created, and their average compression ratios were calculated.}
    \label{fig:2d_runs}
\end{figure}

These results are not surprising, as it is evident that for some mass configurations of the system there are periodic orbits. What is interesting is the fact that a 'general' system is much less 'random' than an actual RNG string. Perhaps certain combinations of masses produce strings that appear more random. To that end, we utilize the Wald-Wolfowitz runs test on strings generated by EM1 for a variety of mass ratios.  In Figure \ref{fig:2d_runs} we can see that the Runs test Z score is very sensitive to mass ratio.  In most cases Z is quite high and collision strings would not be accepted as random binary strings. However, there are a few regions of low Z scores where the collision string would be assessed as coming from a random distribution. 

\begin{figure}[h!]        
    \centering
    \includegraphics[width=1\textwidth]{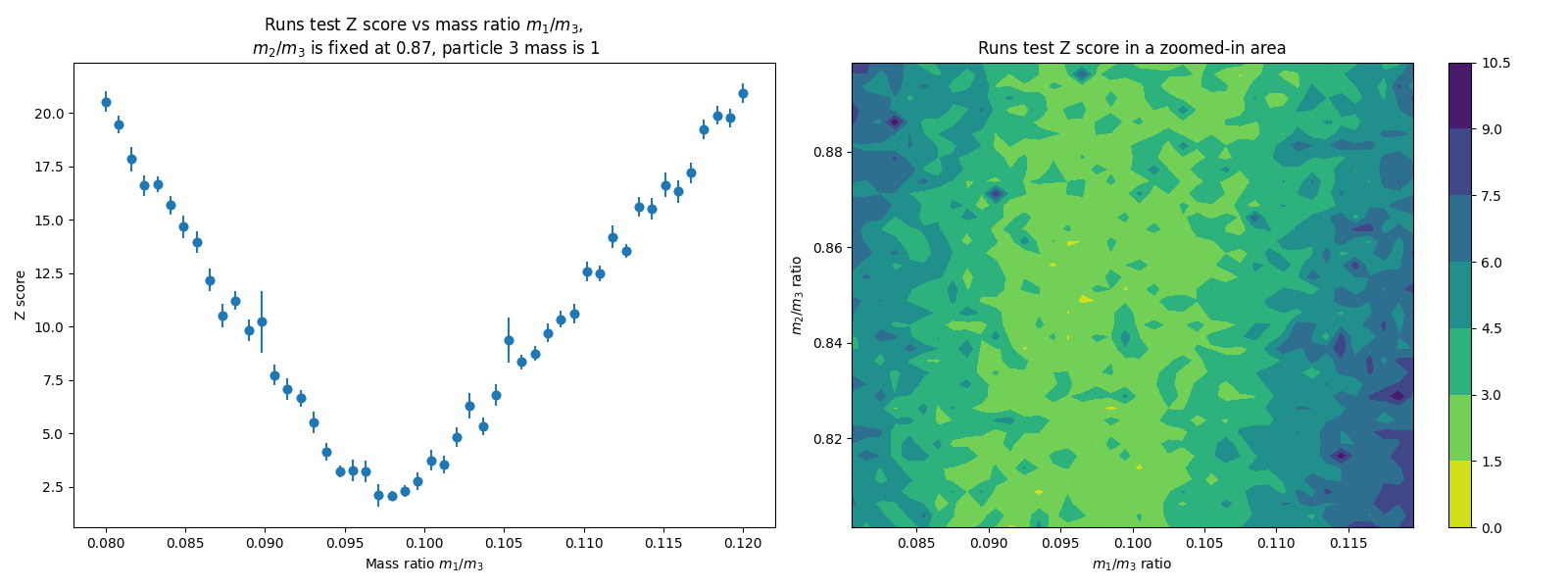}
    \caption{Z scores through an $m_2/m_3$ cut of the mass ratios space are depicted on the left. Minimum value of $Z = 2.08$ occurs for $m_1/m_3 = 0.097$. Although not small enough to confirm statistical randomness with 5\% confidence, it is reasonable to assume that the surrounding area might contain some such mass ratios. Data were sampled from 50 mass ratios; 40 experiments with randomized initial conditions were conducted for each mass ratio. \\
    A zoomed in area of the runs test Z score for differing $m_1/m_3$ and $m_2/m_3$ ratios is depicted on the right. $m_1/m_3$ ratio runs from $0.08$ to $0.12$ while $m_2/m_3$ ratio goes from $0.8$ to $0.9$. For each combination of mass ratios, ten experiments were conducted with randomized initial positions and velocities. Mass ratios were varied from $0.01$ to $1$ for a total of 20 different mass ratios. No intrinsic structure in the distribution of the scores is readily visible.}
    \label{fig:runs1d_andzoom}
\end{figure}

This region of low Z scores was further investigated (Figure \ref{fig:runs1d_andzoom}), and mass ratios that produce a 95\% confidence of randomness were found. There are slight variations in the final results due to the inherent chaotic properties of the system. Different initial starting conditions (positions, velocities) will give out similar, but slightly different results. However there is a clear structure to the region. To emphasize this, a cut was made along the ratio $m_2/m_3 = 0.87$. The minimum average Z score was found to be $2.08$, which is slightly above the 5\% confidence value of 1.96. 

\begin{figure}[h!]
    \centering
    \includegraphics[width=1\textwidth]{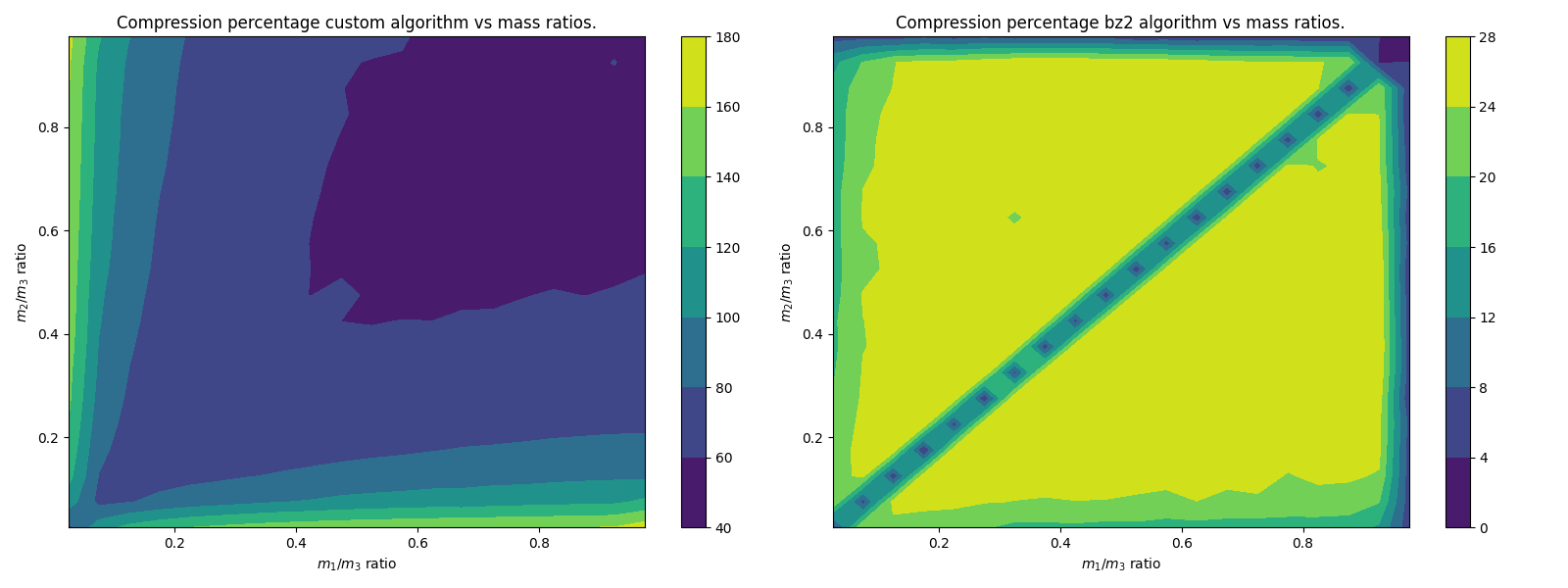}
    \caption{Custom compression algorithm and bz2 algorithm compression results in the plane of mass ratios. The custom compression algorithm reproduces results similar to the runs test, while the bz2 algorithm does not. Bz2 algorithm compresses the strings much better regardless of the mass ratios, and the results do not look similar to the results obtained for RNG strings.
    For a total of 20 equally spaced datapoints, mass ratios were varied from $0.01$ to $1$. Ten strings of length $100,000$ were produced for each mass ratio combination, and their compression ratios were averaged.}
    \label{fig:customandbz2}
\end{figure}

It is interesting to compare these randomness test results with the results of compression algorithms, Figure \ref{fig:customandbz2}. The custom compression algorithm produces the same results as the Z score of the runs test does. A compression percentage of 100$\%$ is obtained in a similar area of the mass ratios. However, the bz2 algorithm does not conform to this result. The compression percentage is lower than the one found in Figure \ref{fig:unified_comparison} for random strings. It does not increase anywhere in the mass ratio space and only gets smaller in certain areas. The bz2 algorithm probably exploits correlations within the string generation mechanism, correlations which the runs test cannot detect. What is interesting is that the areas in which the custom algorithm performs relatively better are the areas in which the bz2 algorithm performs relatively worse, and vice versa.

While none of this is definite proof of the ergodicity of the system it does tell us which mass ratios are relatively better at exploring the phase space of the whole system.
 
\section{Conclusion}
We investigated the relationship between information content and entropy for a dynamical system of three particles on a ring. The method involves the creation of binary  strings representing on the particles' collision history. These collision generated strings were compared to the Random Number Generator strings. This comparison gives us a 'benchmark' against which we can test all our results.

We started with the inspection of information theory entropy, showing that the Shannon entropy is the same for random and EM1 strings, but smaller for EM2 strings. The convergence with run length is noticably faster for the random strings.  Since EM1 and EM2 encode the same information, it implies that the Shannon entropy is not a good measure of information content, presumably because it misses long-range correlations.

Quasi-randomness was tested using the  Wald-Wolfowitz runs test. It was found that mass ratios $\frac{m_1}{m_3} \in [0.1, 0.2]$ and $\frac{m_2}{m_3} \in [0.4, 1]$ produce the lowest runs test scores and are the hardest to distinguish from true randomness. Further exploration of this area of interest produces some deterministic strings that pass the randomness test with $95\%$ confidence. 

We then applied various compression algorithms on differing string lengths. As expected, RNG strings do not get compressed at all. The custom compression algorithm achieved some compression, essentially independent of original string length, but significantly different between EM1 and EM2. On the other hand, bz2 and zlib become more effective at compression as the string length grows. This means that an "entropy" based on bz2/zlib compression efficiency is {\it non-extensive}. This non-extensivity is of a different nature from Tsallis or Renyi entropies which involve different ways of combining p$_i$ - which here are simply $\frac{1}{2}$. Rather, "zlib-entropy non-extensivity" arises because of long-period repetitions in the string which require a different definition of the phase space to account for the fact that each bit depends on the previous one. 

Because the trajectories are deterministic, the actual information content is just the initial conditions, thus asymptotic 100\% compression is in principle possible at the limit of large enough strings. The compression by bz2 appears to be approaching 100\% as string length goes to infinity. 

Since the entropy of both collision generated strings and RNG strings are 1 bit, the non-randomness of the collision generated strings must come from the distribution of the digits within the string and not because of the unequal amounts of digits in the string. bz2 compression uses the Burrows–Wheeler transform and zlib uses Lempel-Ziv, both of which utilize repeated sequences of any length. It implies that compression is achieved due to the existence of patterns at all length scales, a hidden fractal nature emerging from the chaotic dynamics.  

This highlights a stark contrast between our deterministic system and an equivalent stochastic system sampled in the same constant energy, length and momentum ensemble. Properties such as pressure and temperature can be measured directly from momentum transfer and kinetic energy; they can also be measured by phase space integrals assuming ergodicity, and the results are the same.  By contrast, the information content of our trajectory - defined by compression of the collision sequence - is asymptotically zero, whereas a stochastic system has an extensive information content. Our use of the compression entropy gives a quantitative way to show how this chaotic system differs from an ergodic one. Like the FPU \cite{fermi1955studies} paradox, we show that strict ergodicity is a sufficient, but not necessary condition to extract thermodynamic properties from a system.

Overall, we have shown that the deterministic, non-ergodic, three particle on a ring system samples phase space in a way that maximises the entropy, and produces the same energy and momentum distributions as those obtained by phase space integration\cite{ackland1993,Glashow1997,Cox2000}.  The Shannon entropy is equivalent to the thermodynamic entropy from phase space integration, despite the fact that the information content of the strings is low, as demonstrated by the high compressibility.  By contrast, although the bz2 algorithm demonstrates the low information content, it is non-extensive and therefore neither equivalent to thermodynamic entropy nor useful for deriving thermodynamic properties.

Compression algorithms are an attractive and simple way to estimate\cite{tavakoli1993entropy} information content in binary strings, however this information content cannot, in general, be related to a thermodynamic entropy.


\printbibliography[heading=bibintoc, title={References}]

\appendix

\section{Code}
All of the code, scripts and plots used in the project can be found in \href{https://github.com/MatejVe/SHP}{this GitHub repository} \href{https://github.com/MatejVe/SHP}{https://github.com/MatejVe/SHP}.

\section{Python's byte objects}
The final conversion from Python binary strings to the underlying register byte format was one of the project's more severe issues. It is impossible to achieve the desired effects by just compressing a given string, like "0001 0011". Python will encode each character into its ASCII (or UTF-8, or other some encoding scheme) representation and compress those numbers rather than interpreting this string as a single byte. Because ASCII "0" is $30_{16}$ and ASCII "1" is $31_{16}$, the given string will appear very differently in the actual register memory and use up 8 bytes of memory rather than the expected 1: $30_{16}$ $30_{16}$ $31_{16}$ $30_{16}$ $30_{16}$ $31_{16}$ $31_{16}$. This memory representation can be significantly compressed by using the bz2 or zlib compression methods. The string representation must be changed into an in-memory binary number for this reason.
\begin{lstlisting}[language=Python]
    def bitstring_to_bytes(s):
        v = int(s, 2)
        b = bytearray()
        while v:
            b.append(v & 0xff)
            v >>= 8
        return bytes(b[::-1])
\end{lstlisting}
Using a built-in function in Python, we can also create a more condensed version of this function.
\begin{lstlisting}[language=Python]
    def bitstring_to_bytes_compact(s):
        return int(s, 2).to_bytes(-(-len(s) // 8), byteorder='big')
\end{lstlisting}

\section*{Acknowledgement}
For the purpose of open access, the author has applied a Creative Commons Attribution (CC BY) licence to any Author Accepted Manuscript version arising from this submission.
GJA would like to acknowledge the support of the European Research Council (ERC) Grant "Hecate" reference No. 695527. 
The data that support the findings of this study are openly available at the following URL/DOI: https://zenodo.org/badge/latestdoi/451989841

\vspace*{-5mm}

\end{document}